\documentclass[aps,twocolumn,superscriptaddress,showpacs]{revtex4-1}

\usepackage{graphicx}
\usepackage{dcolumn}
\usepackage{bm}
\usepackage{amsmath}
\usepackage{color}
\usepackage{ulem}
\usepackage{threeparttable}
\usepackage{enumerate}
\usepackage{paralist}

\begin{document}

\title{Ising-type quantum spin liquid state in PrMgAl$_{11}$O$_{19}$}

\author{N. Li}
\thanks{These authors contributed equally to this work.}
\affiliation{Anhui Key Laboratory of Magnetic Functional Materials and Devices, Institutes of Physical Science and Information Technology, Anhui University, Hefei, Anhui 230601, People's Republic of China}

\author{A. Rutherford}
\thanks{These authors contributed equally to this work.}
\affiliation{Department of Physics and Astronomy, University of Tennessee, Knoxville, Tennessee 37996, USA}

\author{Y. Y. Wang}
\affiliation{Anhui Key Laboratory of Magnetic Functional Materials and Devices, Institutes of Physical Science and Information Technology, Anhui University, Hefei, Anhui 230601, People's Republic of China}

\author{H. Liang}
\affiliation{Anhui Key Laboratory of Magnetic Functional Materials and Devices, Institutes of Physical Science and Information Technology, Anhui University, Hefei, Anhui 230601, People's Republic of China}

\author{Q. J. Li}
\affiliation{School of Physics and Optoelectronics, Anhui University, Hefei, Anhui 230061, People's Republic of China}

\author{Z. J. Zhang}
\affiliation{School of Information $\&$ Electronic Engineering (Sussex Artificial Intelligence Institute), Zhejiang Gongshang University, Hangzhou, Zhejiang 310018, People's Republic of China}

\author{H. Wang}
\affiliation{Department of Chemistry, Michigan State University, East Lansing, Michigan 48824, United States}

\author{W. Xie}
\affiliation{Department of Chemistry, Michigan State University, East Lansing, Michigan 48824, United States}

\author{H. D. Zhou}
\email{hzhou10@utk.edu}
\affiliation{Department of Physics and Astronomy, University of Tennessee, Knoxville, Tennessee 37996, USA}

\author{X. F. Sun}
\email{xfsun@ahu.edu.cn}
\affiliation{Anhui Key Laboratory of Magnetic Functional Materials and Devices, Institutes of Physical Science and Information Technology, Anhui University, Hefei, Anhui 230601, People's Republic of China}

\date{\today}

\begin{abstract}

We have grown single crystals of PrMgAl$_{11}$O$_{19}$, an ideal triangular-lattice antiferromagnet, and performed magnetic susceptibility, specific heat and thermal conductivity measurements at low temperatures. The main results are as follows: (i) The temperature-dependent susceptibility shows a negligible in-plane response and the isothermal magnetization curves confirm the easy axis along the $c$ axis. (ii) The specific heat measurements reveal the absence of long-range magnetic order down to 60 mK, and the power-law temperature dependence indicates the existence of the gapless magnetic excitations in system. (iii) The ultralow-temperature thermal conductivity exhibits negligibly small residual term ($\kappa_0/T$) and strong spin-phonon scattering effect, suggesting that the spin excitations are also involved. Our results further demonstrate that PrMgAl$_{11}$O$_{19}$ is a rare quantum spin liquid candidate with Ising-like anisotropy.

\end{abstract}


\maketitle

\section{Introduction}

The strong frustration effect in antiferromagnets leads to strong quantum fluctuations that prevent the formation of long-range magnetic order. These systems have rich and exotic quantum magnetisms, quantum phase transition behaviors, and peculiar elementary excitations \cite{PhysToday59}. In fact, the most striking physical state is the quantum spin liquid (QSL), in which the strongly interacted spins do not develop a conventional long-range magnetic order even at the absolute zero-Kelvin temperature limit, but highly correlate with each other due to long-range quantum entanglement. Moreover, QSLs have some special elementary excitations, including spinons, Majorana fermions and topological visons, which are closely related to quantum information, quantum computing and high temperature superconductivity. Therefore, QSL has attracted extensive attentions and become one of the focuses in condensed matter physics \cite{Nature464, RevModPhys89, RevCondens10}.

In 1973, Anderson first proposed the resonating valence bond (RVB) model to describe this novel quantum disorder state based on Heisenberg antiferromagnets with two-dimensional triangular lattices \cite{MaterResBull8}. Since then, many efforts have been devoted to realizing this exotic quantum disorder state in real materials, especially in low-dimensional antiferromagnetically correlated systems with small spin quantum numbers or strong frustration effects. Typical research systems are two-dimensional triangular lattice and kagome lattice, three-dimensional pyrochlore lattice, and more recently, two-dimensional honeycomb lattice. Among them, the triangular-lattice is the most representative system, in which the organic antiferromagnets $\kappa$-(BEDTTTF)$_2$Cu$_2$(CN)$_3$ and EtMe$_3$Sb[Pd(dmit)$_2$]$_2$ with $S =$ 1/2 were identified as QSL candidates \cite{Science372, Science328}. For inorganic materials, some new QSL candidates have also emerged, including YbMgGaO$_4$, $ARCh_2$ ($A =$ alkali or monovalent ions, $R =$ rare earth ions, $Ch =$ chalcogenides), Na$_2$BaCo(PO$_4$)$_2$, Pr$M$Al$_{11}$O$_{19}$ ($M =$ Zn, Mg), NdTa$_7$O$_{19}$, Ba$_6$$R_2$Ti$_4$O$_{17}$ ($R =$ rare earth ions), YbZn$_2$GaO$_5$, NaRuO$_2$, TbInO$_3$ etc \cite{SciRep5, CPL35, PNAS116, JAlloysCompd, JMaterChem, NatMater416, arXiv08937, arXiv20040, NatPhys19, NatPhys15}.

The inevitable disorder effects such as vacancies, impurities, lattice distortion, and stacking faults in real materials often cause many controversies in QSL candidates. The canonical examples including the kagome-lattice system Zn$_3$Cu(OH)$_6$Cl$_2$ and the triangular-lattice ytterbium-based compound YbMgGaO$_4$. For Zn$_3$Cu(OH)$_6$Cl$_2$, it has been confirmed that about 5 $\sim$ 15 \% of magnetic Cu$^{2+}$ ions on the interlayer sites occupied by non-magnetic Zn$^{2+}$ ions, and this structural disorder may affect the magnetic exchange interactions significantly \cite{JAmChemSoc132, RevModPhys88}. Therefore, it is difficult to accurately characterize the real ground sate in theory and experiments. Moreover, unlike the case of Zn$_3$Cu(OH)$_6$Cl$_2$, there is a site mixing between nonmagnetic Mg$^{2+}$ and Ga$^{3+}$ ions in YbMgGaO$_4$, causing the controversy on the ground state and the low-energy magnetic excitations \cite{Nature540, PhysRevLett117, PhysRevLett119}. Consequently, there is an urgent need for an ideal system to investigate the QSL states in a better way.

Recently, it has been reported that Pr$M$Al$_{11}$O$_{19}$ ($M =$ Zn, Mg) is a promising QSL candidate with perfect two-dimensional triangular-lattice with only a weak disorder between nonmagnetic $M^{2+}$ and Al$^{3+}$ ions, which has a negligible effect on the ground state \cite{PhysRevB165143, PhysRevB134428}. Previous studies indicated the absence of long-range magnetic order at ultralow temperatures, and the results of specific heat and inelastic neutron scattering revealed the existence of gapless low-energy excitations, in favor of the possible gapless QSL in Pr$M$Al$_{11}$O$_{19}$ ($M =$ Zn, Mg). However, more experimental evidence by using other techniques are still called for. In this regard, ultralow-temperature thermal conductivity has been proved to be a clean and powerful technique in studying the low-energy magnetic excitations in QSL candidates \cite{Science328, NC4949, NC4216}.

In this work, we have grown high-quality PrMgAl$_{11}$O$_{19}$ single crystal and have systematically studied its magnetic properties, specific heat and thermal conductivity at low temperatures. The results indicate no sign of long-range magnetic order at low temperatures down to several tens of milliKelvin. The temperature-dependent susceptibility and the isothermal magnetization indicate Ising-like anisotropy with easy axis along the $c$ axis. The negative Curie-Weiss temperature obtained from the magnetic susceptibility shows the dominant antiferromagnetic interactions between Pr$^{3+}$ ions. A clear power-law temperature dependence is observed on the low-temperature specific heat, suggesting the existence of gapless spinon excitations. Besides, the main experimental result of present work is that the ultralow temperature thermal conductivity displays a rough $T^2$ dependence and a negligibly small residual term $\kappa_0/T$. It indicates the strong spin-phonon scattering and therefore the existence of spinons at ultralow temperatures. Our results further verify the realization of Ising-like QSL in PrMgAl$_{11}$O$_{19}$.

\section{Experiments}

PrMgAl$_{11}$O$_{19}$ single crystals were grown using the floating zone method. The feed and seed rods for the crystal growth were prepared by solid state reactions. The stoichiometric  mixtures of Pr$_6$O$_{11}$ (pre-dried at 1000 $^{\circ}$C for overnight), MgO, and Al$_2$O$_3$ were ground together and pressed into 6-mm-diameter 60-mm rods under 400-atm hydrostatic pressure and then calcined in air at 1000 $^{\circ}$C for 20 hours and at 1400 $^{\circ}$C for 20 hours, and finally in argon at 1400 $^{\circ}$C for 20 hours with intermediate grindings. The crystal growth was carried out in argon in an IR-heated image furnace equipped with two halogen lamps and double ellipsoidal mirrors with feed and seed rods rotating in opposite directions at 20 rpm during crystal growth at a rate of 2.0 mm/hour. The obtained PrMgAl$_{11}$O$_{19}$ single crystal with size of several centimeters is shown in Fig. 1(b). The Laue back diffraction measurement was used to determine the crystalline orientation. The single crystal X-ray diffraction (SCXRD) measurement was performed using a XtalLAB Synergy, Dualflex, Hypix single crystal X-ray diffractometer with Mo K$_{\alpha}$ radiation ($\lambda$ = 0.71073 \AA). The structure was solved and refined using the Bruker SHELXTL Software Package.

The dc magnetic susceptibility between 2 and 300 K were measured using the SQUID-VSM (Quantum Design Magnetic Property Measurement System, MPMS) in magnetic field up to 7 T. The dc magnetization up to 14 T was measured using a VSM equipped with the Physical Property Measurement System (PPMS) (DynaCool, Quantum Design). The specific heat was measured by relaxation technique using PPMS with a dilution insert. The same sample was used for both magnetic properties and specific heat measurements. The PrMgAl$_{11}$O$_{19}$ single crystal for thermal conductivity measurements was cut precisely along the $a$ axis with dimension of 2.5 $\times$ 0.64 $\times$ 0.16 mm$^3$ after being oriented by using the X-ray Laue system. The thermal conductivity was measured by using ``one heater, two thermometers" technique in a dilution refrigerator (70 mK to 1 K) and a $^3$He refrigerator (0.3 to 30 K) equipped with a 14 T magnet \cite{NC4949, NC4216, PhysRevB184423}. The heat current was along the longest direction ($a$ axis) and the external magnetic fields were applied along either the $a$ or $c$ axis. It should be pointed out that PrMgAl$_{11}$O$_{19}$ single crystal is so easily cleaved along the $ab$ plane that it is not possible to cut a long-bar shaped sample along the $c$ axis.

\section{Results and Discussion}

\subsection{Crystal structure}

\begin{table*}[htbp]
	\caption{The crystal structure and refinement for PrMgAl$_{11}$O$_{19}$ at 300 K.} 
	\centering 
	\begin{tabular}{c c} 
		\hline\hline 
	    $\textbf{Chemical}$ $\mathbf{Formula}$ & $\textbf{PrMgAl$_{11}$O$_{19}$}$ \\ 
		\hline 
        Formula weight & 766.00 g/mol \\ 
        Space Group & $P6_3/mmc$ \\
		Unit cell dimensions & $a =$ 5.58865(6) \AA \\
        ~\ & $c =$ 21.9115(4) \AA \\
	    Volume & 592.68(2) \AA$^3$ \\
		Density(calculated) & 4.292 g/cm$^3$ \\
		Extinction coefficient & 0.010(1) \\
		Absorption coefficient & 5.130 mm$^{-1}$ \\
        F(000) & 732 \\
        2$\theta$ range & 7.44 to 80.84$^{\circ}$ \\
        Total Reflections & 17535 \\
        Independent reflections & 782 [$R\rm_{int} =$ 0.0356] \\
        Refinement method & Full-matrix least-squares on F$^2$ \\
        Data/restraints/parameters & 782 / 0 / 42 \\
        Final R indices & $R_1$ ($I >$ 2$\sigma$($I$)) = 0.0341; $wR_2$ ($I >$ 2$\sigma$($I$)) = 0.0838 \\
        ~\ & $R_1$(all) = 0.0346; $wR_2$ (all) = 0.0841 \\
        Largest diff. peak and hole & +6.554 e/\AA$^{-3}$ and -3.504 e/\AA$^{-3}$ \\
        R.M.S. deviation from mean & 0.264 e/\AA$^{-3}$ \\
        Goodness-of-fit on F$^2$ & 1.068 \\

		\hline 
	\end{tabular}
	\label{table:nuclear} 
\end{table*}

\begin{table*}[htbp]
	\caption{Atomic coordinates and equivalent isotropic atomic displacement parameters (\AA$^2$). (U$_{eq}$ is defined as one third of the trace of the orthogonalized U$_{ij}$ tensor.)} 
	\centering 
	\begin{tabular}{c c c c c c c} 
		\hline\hline 
		$\textbf{PrMgAl$_{11}$O$_{19}$}$ & $\textbf{Wyck.}$ & $\textbf{x}$ & $\textbf{y}$ & $\textbf{z}$ & $\textbf{Occ.}$ & $\textbf{U$_\text{eq}$}$ \\ [0.5ex] 
		\hline 
		Pr & 2$d$ & 1/3 & 2/3 & 3/4 & 1 & 0.0052(1) \\ 
		Al$_1$ & 2$a$ & 0 & 0 & 0 & 1 & 0.0011(2) \\
		Al$_2$ & 2$b$ & 0 & 0 & 1/4 & 1 & 0.0011(3) \\
        Al$_3$ & 4$f$ & 1/3 & 2/3 & 0.18988(2) & 1 & 0.0011(2) \\
        Al$_4$/Mg & 4$f$ & 1/3 & 2/3 & 0.02721(3) & 0.5/0.5 & 0.0008(1) \\
        Al$_5$ & 12$k$ & 0.16749(3) & 0.33501(6) & 0.60834(2) & 1 & 0.0009(1) \\
		O$_1$ & 4$e$ & 0 & 0 & 0.15151(6) & 1 & 0.0014(3) \\
		O$_2$ & 4$f$ & 1/3 & 2/3 & 0.55803(6) & 1 & 0.0067(9) \\
        O$_3$ & 6$h$ & 0.1812(1) & 0.3624(1) & 1/4 & 1 & 0.0017(3) \\
        O$_4$ & 12$k$ & 0.5054(1) & 0.0109(1) & 0.1516(1) & 1 & 0.0015(3) \\
        O$_5$ & 12$k$ & 0.1522(1) & 0.3045(1) & 0.0537(1) & 1 & 0.0037(4) \\
		\hline 
	\end{tabular}
	\label{table:nuclear2} 
\end{table*}

\begin{figure}
\includegraphics[clip,width=8.5cm]{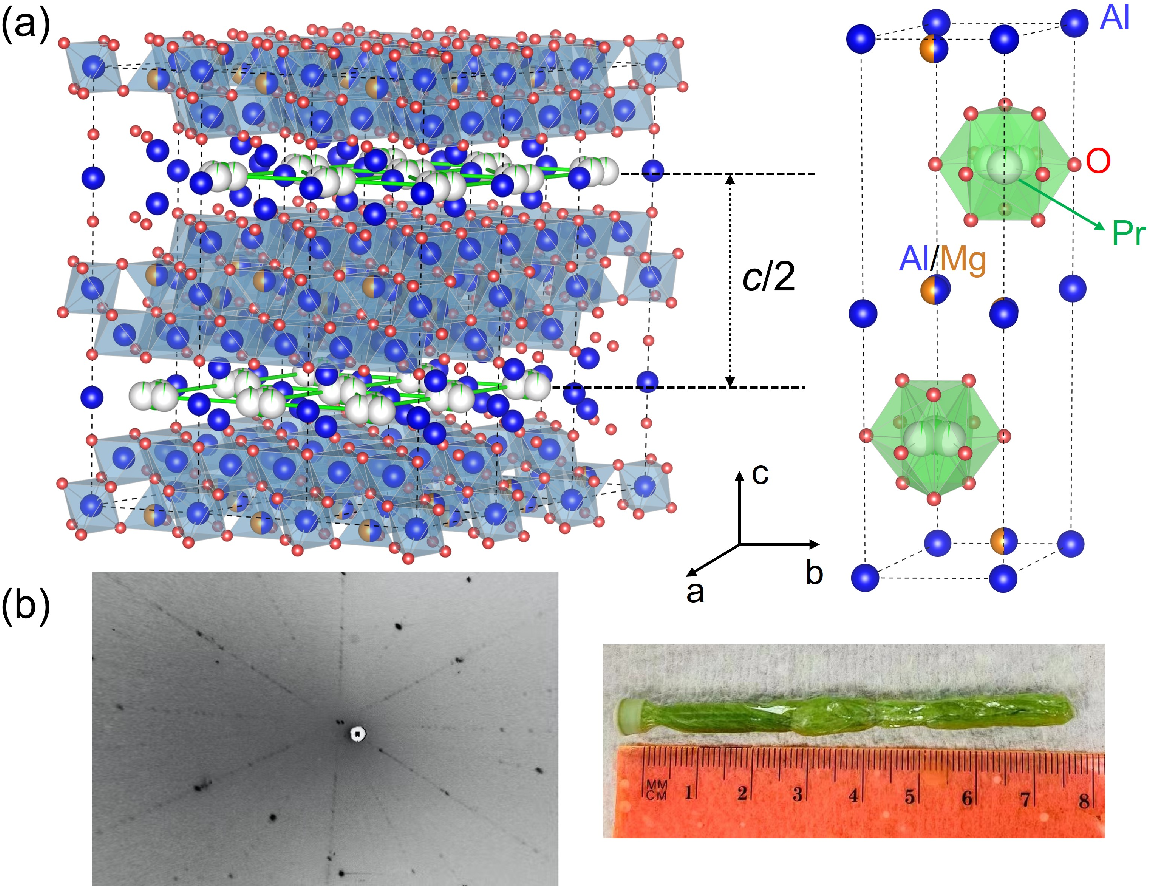}
\caption{(a) The schematic crystal structure of PrMgAl$_{11}$O$_{19}$, where Pr$^{3+}$ ions form into an ideal triangular layer in the $ab$ plane and these triangular layers are stacked along the $c$ axis with the nearest-neighbor interlayer distance of 10.9558 \AA. (b) The X-ray Laue back diffraction photo for the $ab$ plane and the photograph of PrMgAl$_{11}$O$_{19}$ single crystal grown by using the optical floating-zone technique.}
\label{S}
\end{figure}

The crystalline structure of as-grown PrMgAl$_{11}$O$_{19}$ crystal is determined by the SCXRD data refinement with a $P6_3/mmc$ space group. The obtained lattice parameters and atomic coordinates are summarized in Tables I and II. The schematic crystal structure of PrMgAl$_{11}$O$_{19}$ is shown in Fig. 1(a). The magnetic Pr$^{3+}$ ion is coordinated within a PrO$_{12}$ tetrakaidecahedron and form an ideal triangular-lattice layer with the nearest-neighbor Pr-Pr intra-layer distance of $a =$ 5.58865(6) \AA. These triangular-lattice layers are stacked along the $c$ axis with $c =$ 21.9115(4) \AA, separated by nonmagnetic AlO$_6$ octahedra and (Mg, Al)O$_4$ tetrahedra, which indicates a quasi-two-dimensional structural characteristic of PrMgAl$_{11}$O$_{19}$. Due to the large ion radius difference between magnetic (Pr$^{3+}$) and nonmagnetic (Mg$^{2+}$/Al$^{3+}$) ions, the site mixing disorder is eliminated. The obtained green crystal is easy to cleave along the $ab$ plane and this orientation was confirmed by Laue back diffraction pattern (Fig. 1(b)).

\subsection{Magnetic Properties}

\begin{figure}
\includegraphics[clip,width=8.5cm]{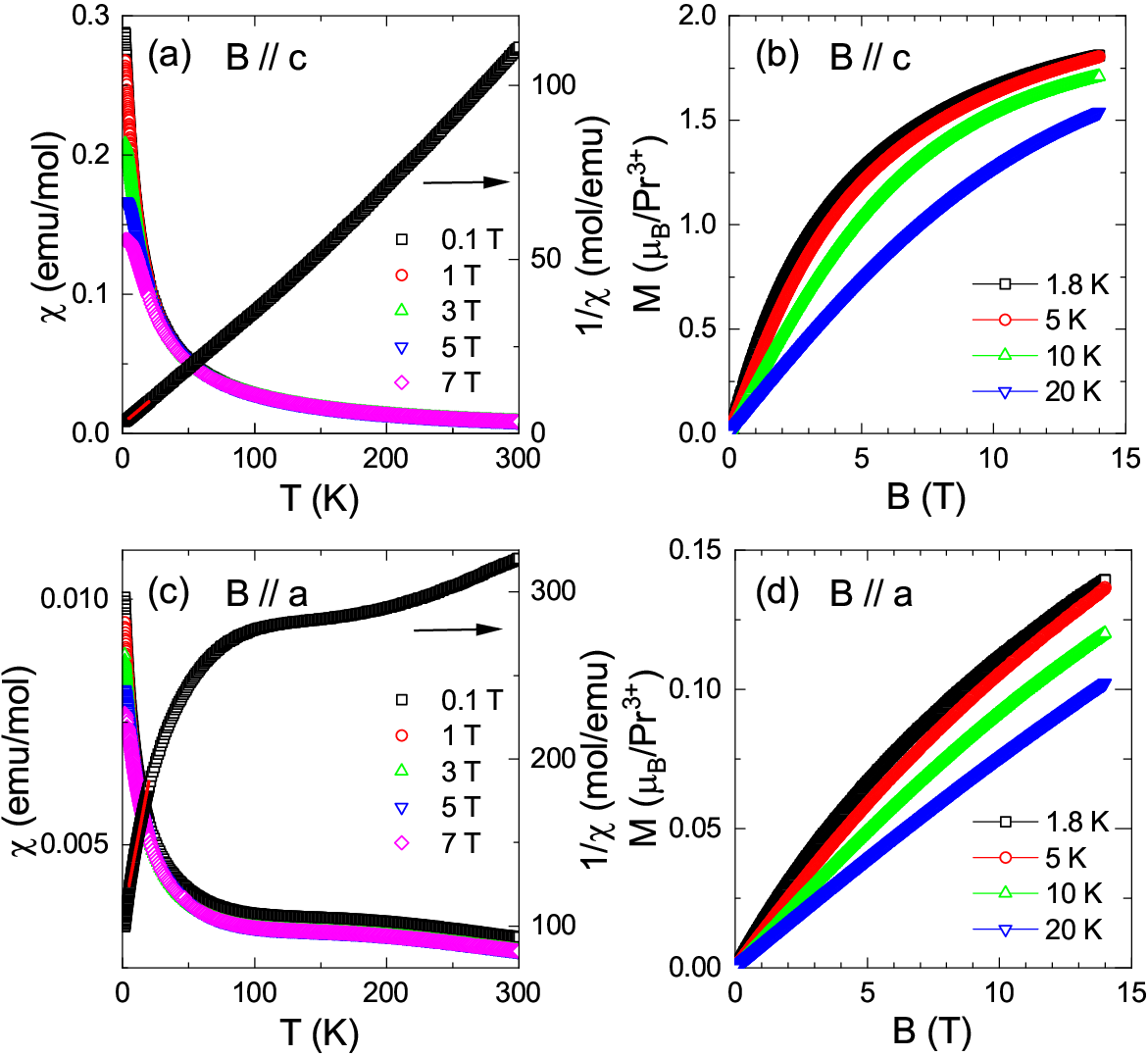}
\caption{(a, c) Temperature dependence of dc magnetic susceptibility and the inverse magnetic susceptibility with external magnetic fields along the $c$ or $a$ axis. The solid red lines indicate the CW fitting at 2 $\leq$ $T$ $\leq$ 20 K under $B =$ 0.1 T. (b, d) Magnetic field dependence of the magnetization at selected temperatures with $B \parallel c$ and $B \parallel a$, respectively.}
\label{MH}
\end{figure}

Magnetic susceptibility $\chi(T)$ measurements were performed from 2 to 300 K under different magnetic fields along the $c$ or $a$ axis, as shown in Figs. 2(a) and 2(c). The absence of any anomaly in $\chi(T)$ curves indicates that the Pr$^{3+}$ moments do not undergo a long range magnetic ordering down to 2 K. The absence of splitting of $\chi(T)$ between zero field cooling (ZFC) and field cooling (FC) under 0.1 T precludes spin freezing. Due to the presence of crystal electric field (CEF) effect, the $\chi(T)$ curves deviate the Curie-Weiss (CW) behavior at the intermediate temperature. The low-temperature (2 -- 20 K) CW fittings to the 0.1 T data yield $\theta\rm_{CW} = -$ 8.44 K, $\mu_{\text{eff}}$ = 5.0 $\mu_{\text{B}}$/Pr and $\theta\rm_{CW} = -$ 24.85 K, $\mu_{\text{eff}}$ = 1.39 $\mu_{\text{B}}$/Pr for $B \parallel c$ and $B \parallel a$, respectively. The negative $\theta\rm_{CW}$ reveals dominant antiferromagnetic exchange interaction between Pr$^{3+}$ ions. For $B \parallel c$, the magnetic susceptibility at high temperatures are almost independent on the external magnetic field, while the susceptibility at low temperatures gradually decreases with increasing magnetic field. In contrast, for $B \parallel a$, the magnetic susceptibility is two orders of magnitude smaller compared with $B \parallel c$, and the magnetic susceptibility gradually reduces with increasing magnetic field, which suggests the Ising-like anisotropy in PrMgAl$_{11}$O$_{19}$ \cite{PhysRevB054435}.

The isothermal field-dependent magnetization curves $M(B)$ at selected temperatures are shown in Figs. 2(b) and 2(d) for $B \parallel c$ and $B \parallel a$, respectively. At these temperatures, the $M(B)$ curves exhibit nonlinear field dependence without magnetic saturation up to 14 T for both directions, and the maximum magnetization values are about 1.79 $\mu_{\text{B}}$/Pr$^{3+}$ and 0.14 $\mu_{\text{B}}$/Pr$^{3+}$, respectively, much smaller than the theoretical saturation value of 3.58 $\mu_{\text{B}}$/Pr$^{3+}$ for free ions. These data further indicate large spin anisotropy with easy axis along the $c$ axis.

\begin{figure}
\includegraphics[clip,width=6cm]{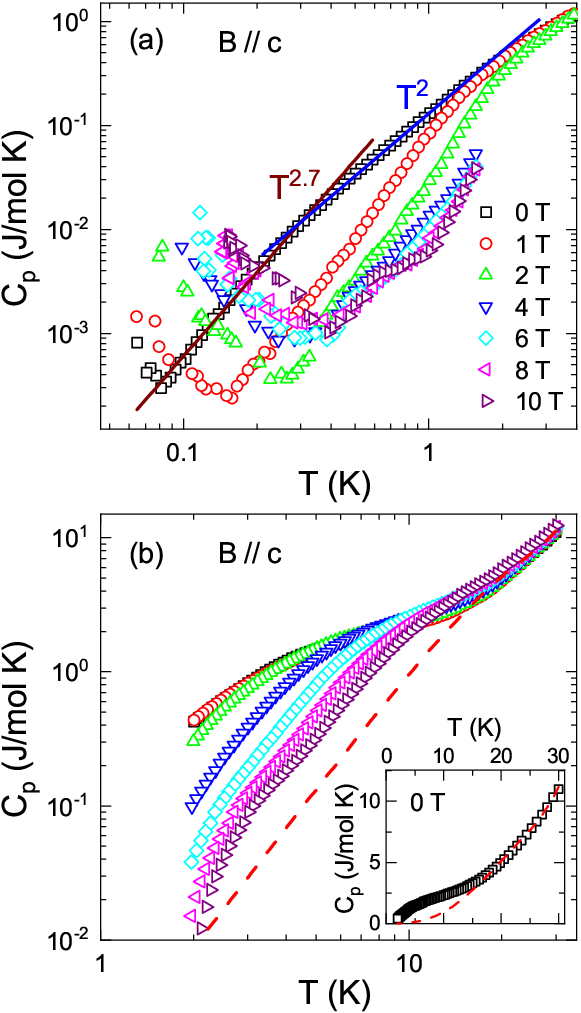}
\caption{(a) Ultralow-temperature specific heat $C_p$($T$) of PrMgAl$_{11}$O$_{19}$ single crystal under different magnetic fields along the $c$ axis. The solid lines indicate the $T^{2.7}$ and $T^2$ dependence, respectively. (b) The temperature dependence of specific heat with external magnetic fields along the $c$ axis at 2 $< T <$ 30 K. Inset: the zero-field specific heat data. The dashed lines show the the fitting to the high-$T$ data by using the formula of phonon specific heat $C = \beta T^3 + \beta_5 T^5 + \beta_7 T^7$.}
\label{Cp}
\end{figure}

To detect the low-energy excitations of PrMgAl$_{11}$O$_{19}$ single crystal, we perform the specific heat measurements from 60 mK to 30 K with $B \parallel c$, and the temperature dependence of specific heat $C_p$($T$) curves are shown in Fig. 3. It can be seen that there is no sharp $\lambda$-shape anomaly down to several tens of milliKelvin, indicating the absence of long-range magnetic order of Pr$^{3+}$ ions despite of the presence of a large AFM interaction ($\theta\rm_{CW} = -$ 8.44 K), which may imply the possible realization of the QSL state. Noted that the lowest-temperature upturns of $C_p$($T$) is related to the nuclear Schottky anomaly, which shifts to higher temperature with increasing magnetic field. Moreover, the zero-field $C_p$($T$) curve exhibits a hump around 4 K indicative of low-lying crystal field excitations, which is commonly observed in some rare-earth-based frustrated triangular-lattice antiferromagnets \cite{arXiv2108, PhysRevM114413, PhysRevB214408, PhysRevBL121109}. The hump is gradually suppressed and shifts to higher temperatures with increasing external magnetic field, as shown in Fig. 3(b).

With increasing magnetic field, the low-temperature specific heat is significantly decreased, besides the lowest-temperature upturn of the nuclear Schottky term. This indicates that at zero field the low-temperature specific heat is dominated by the magnetic excitations while the phonon term is negligibly small. As we can seen that the specific heat obeys the $T^{2.7}$ and $T^2$ power-law temperature dependence at 80 mK $< T <$ 300 mK and 300 mK $< T <$ 2 K, respectively. Although the power exponent is sensitive to the fitting temperature range, which is similar to kagome compound ZnCu$_3$(OH)$_6$Cl$_2$, it still may evidence the gapless magnetic excitations \cite{PhysRevLett107204}. In fact, this power-law temperature dependence has already been reported in some QSL candidates as one of the experimental hallmarks for gapless low-energy magnetic excitations, such as the triangular-lattice compound YbMgGaO$_4$ and YbZn$_2$GaO$_5$ \cite{SciRep5, arXiv20040}, the kagome-lattice compound YCu$_3$(OH)$_{6+x}$Br$_{3-x}$ ($x \approx$ 0.5) and ZnCu$_3$(OH)$_6$Cl$_2$ \cite{PhysRevBL121109, NatPhys435, PhysRevLett107204}. In particular, the sister material PrZnAl$_{11}$O$_{19}$ also show exhibits a $T^{1.897}$ behavior of zero-field magnetic specific heat at $\sim$ 0.3 -- 3 K and some slope change at lower temperatures (the measurement was done only down to 200 mK) \cite{PhysRevB134428}. Furthermore, the recent study on PrMgAl$_{11}$O$_{19}$ also revealed a $T^2$ behavior of zero-field magnetic specific heat at $\sim$ 0.3 -- 1 K, as well as a different temperature dependence at lower temperatures \cite{PhysRevB165143}. Under 1 T or 2 T fields the low-temperature specific heat have a similar dependence to that in zero field. At higher fields, the low-temperature Schottky term enhances and strongly affects the temperature dependence of specific heat.

We have done quantitative analysis on phonon contribution of the specific heat. It is known that in the temperature range 0.02 $< T / \Theta_D <$ 0.1 ($\Theta_D$ is the Debye temperature), one can use the low-frequency expansion of the Debye function, $C = \beta T^3 + \beta_5 T^5+\beta_7 T^7$, where $\beta$, $\beta_5$ and $\beta_7$ are temperature-independent coefficients \cite{Tari, GFO}. As shown in Fig. 3(b), the fitting to the zero-field data at relatively high temperatures yields parameters $\beta = 1.09 \times 10^{-3}$ J/K$^4$mol, $\beta_5 = -1.43 \times 10^{-6}$ J/K$^6$mol and $\beta_7 = 7.60 \times 10^{-10}$ J/K$^8$mol. At very low temperatures, the $T^5$- and $T^7$- terms are negligible and the phonon specific heat shows a well-known $T^3$ dependence with the coefficient of $\beta$. Note that this fitting result also indicates that the phonon specific heat is negligibly small at low temperatures.

\subsection{Thermal conductivity}

\begin{figure}
\includegraphics[clip,width=8.5cm]{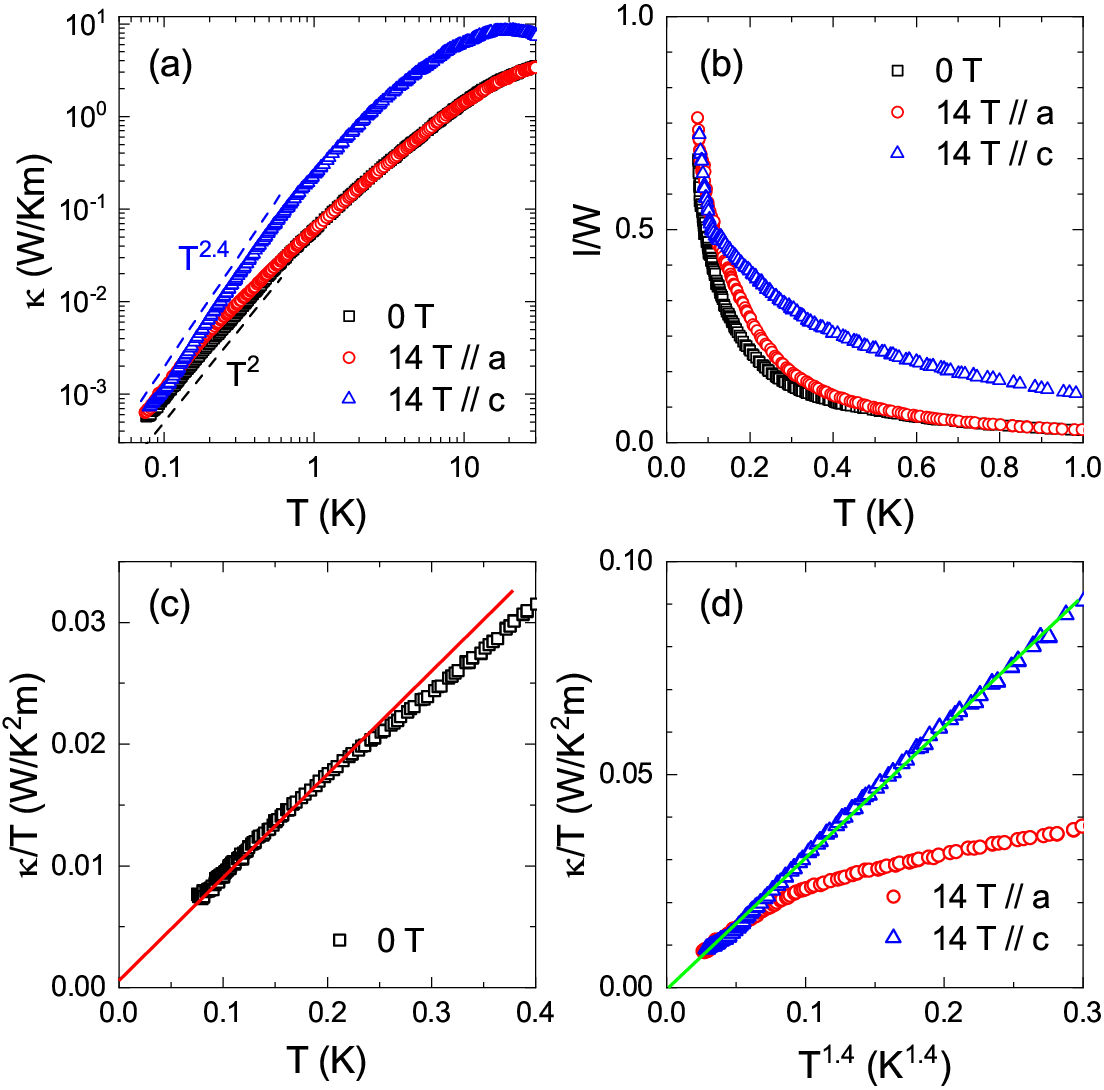}
\caption{(a) Temperature dependence of the $a$-axis thermal conductivity $\kappa(T)$ of PrMgAl$_{11}$O$_{19}$ single crystal under zero field and 14 T magnetic field applied along either the $a$ or $c$ axis, the dashed lines indicate the $T^2$- and $T^{2.4}$-behavior at zero field and 14 T magnetic field applied along the $c$ axis, respectively. (b) Temperature dependence of the phonon mean free path $l$ divided by the averaged sample width $W$. (c) The $\kappa/T$ plotted as a function of $T$ for ultralow temperature data at zero field. The solid line is the linear fitting of the data at $T <$ 200 mK by using the formula $\kappa/T =$ $\kappa_0/T$ + $bT$, with a negligibly small residual thermal conductivity $\kappa_0/T$. (d) The $\kappa$/$T$ plotted as a function of $T^{1.4}$ for $T <$ 420 mK at 14 field along the $a$ or $c$ axis. The solid line is the linear fitting to the $c$-axis-field data with zero $\kappa_0/T$. The $a$-axis-field data also point to a vanishing $\kappa_0/T$.}
\label{kaT}
\end{figure}

Figure 4 shows the low-temperature thermal conductivity of PrMgAl$_{11}$O$_{19}$ single crystal in zero and in 14 T magnetic field along the $a$ or $c$ axis. At zero field, the $\kappa(T)$ curve exhibits a rough $T^2$-dependence at 70 mK $< T <$ 600 mK, as shown in Fig. 4(a), which significantly deviates from the standard $T^3$ behavior of phonon thermal conductivity at very low temperatures. It usually suggests that the phonons are strongly scattered with magnetic excitations \cite{Oxford}. When a 14 T magnetic field is applied along the $a$ axis, the $\kappa$ increases slightly at low temperature (70 mK $< T <$ 1 K) and almost overlaps with the $\kappa$ for $B =$ 0 T at higher temperatures. In contrast, the $\kappa$ is significantly enhanced and displays a stronger $T^{2.4}$-dependence for 14 T $\parallel c$. This indicates that the magnetic scattering on phonon is strongly smeared out under the 14 T magnetic field along the $c$ axis. It can be seen that the different responses of the $\kappa$ for $B \parallel c$ and $B \parallel a$ further verify the Ising-like properties of PrMgAl$_{11}$O$_{19}$. That is, even such a quite strong field along the $a$ axis can hardly affect the spin system or spin excitations, whereas the 14 T along the $c$ axis seems to strongly suppress the QSL phase and its spinon excitations.

It is possible to estimate the mean free path of phonons at low temperatures if the $\kappa$ is mainly contributed by the phonons. The phononic thermal conductivity can be expressed by the kinetic formula $\kappa_{ph} = \frac{1}{3}Cv_pl$ \cite{Oxford}, where $C = \beta T^3$ is phonon specific heat at low temperatures, $v_p$ is the average velocity and $l$ is the mean free path of phonon. Here $\beta = 1.09 \times 10^{-3}$ J/K$^4$mol is obtained from the zero-field specific-heat data and $v_p$ = 2710 m/s can be estimated from Deybe temperature $\Theta_D$ using the relations $\beta = \frac{12\pi^4}{5} \frac{Rs}{\Theta_D^3}$ and $\Theta_D = \frac{\hbar v_p}{k_B} (\frac{6\pi^2 Ns}{V})^\frac{1}{3}$ \cite{Tari}, where $N$ is the number of molecules per mole and each molecule comprises $s$ atoms, $V$ is the volume of crystal and $R$ is the universal gas constant. Figure 4(b) shows the ratio of calculated $l$  to the averaged sample width $W = 2\sqrt{A/\pi}$ = 0.367 mm \cite{Oxford, PhysRev176501}, where $A$ is the area of cross section. It can be seen that the ratio $l / W$ increases with lowering temperature and becomes close to one at the lowest temperature, which means that the boundary scattering limit is nearly established at such low temperatures. It should be noted that these phonon mean free paths would be over-estimated if there were sizeable contribution to $\kappa$ from other heat carriers.

Ultralow-temperature thermal conductivity is an effective tool to detect the characteristics of magnetic excitations in the system. Usually, one can use the formula $\kappa/T =$ $\kappa_0/T$ + $bT^{\alpha-1}$ to fit the thermal conductivity of insulator, where $\kappa_0/T$ represents a constant contribution from gapless fermionic excitations and $bT^{\alpha}$ represents the phonon thermal conductivity with the exponent $\alpha =$ 2 $\sim$ 3 \cite{PhysRev176501, PhysRevB184423}. For the zero-field data plotted as $\kappa/T$ vs $T$, as shown in Fig. 4(c), the linear fitting at $T <$ 200 mK gives an almost zero residual thermal conductivity, $\kappa_0/T$. Moreover, the larger $\kappa$ in high fields indicates that the phonons are rather strongly scattered by magnetic excitations, which are gapped out in high magnetic fields. Therefore, the absence of magnetic excitations in the thermal conductivity result at zero magnetic field is due to the spin-phonon scattering that not only weakens the phonon transport but also prevents the spinon transport. In this regard, it seems to be consistent with the recent neutron scattering measurements of PrMgAl$_{11}$O$_{19}$ that revealed the gapless spinon excitations \cite{PhysRevB165143}.

For the data with 14 T $\parallel c$, since the temperature dependence is close to $T^{2.4}$, we can have a rather good linear fitting in the $\kappa/T$ vs $T^{1.4}$ plot at $T <$ 420 mK, which also gives a zero residual term, as shown in Fig. 4(d). For the low-temperature data with 14 T $\parallel a$, which are slightly larger than those at zero field, they display some clear change of the temperature dependence at about 200 mK. Nevertheless, the lowest-temperature data with 14 T $\parallel a$ are very comparable to those at 14 T $\parallel c$ and therefore also point to a zero $\kappa_0/T$, as shown in Fig. 4(d). The absence of fermions contribution to the $\kappa$ with applying strong magnetic field is always the case for QSL candidates, due to the suppression of QSL phase by magnetic field.

\begin{figure}
\includegraphics[clip,width=6cm]{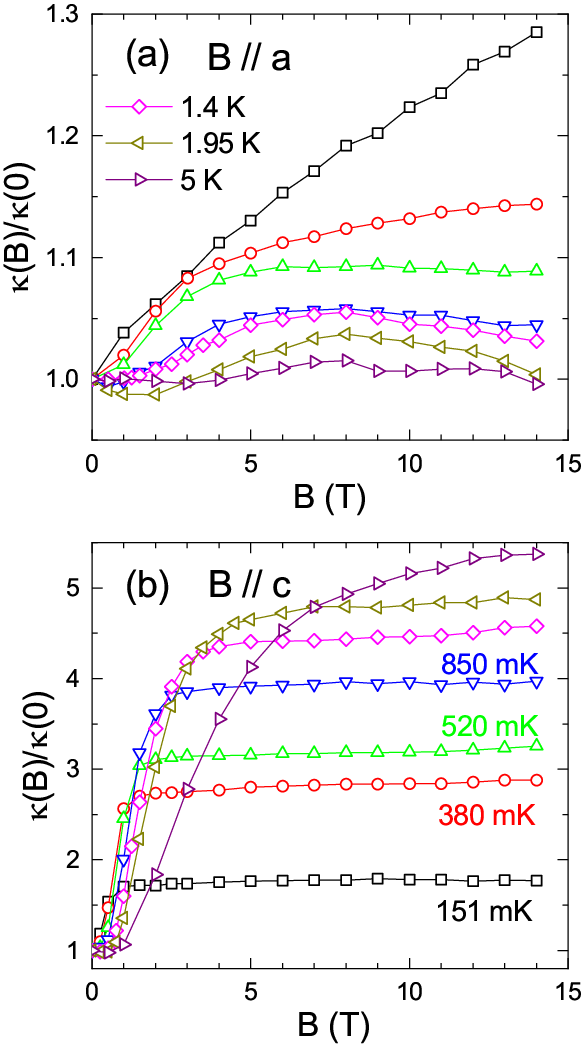}
\caption{The magnetic field dependence of thermal conductivity at different temperatures with external magnetic field along the $a$ and $c$ axis, respectively.}
\label{kaH}
\end{figure}

Figure 5 shows the magnetic field dependence of thermal conductivity at different temperatures with external magnetic field along the $a$ or $c$ axis. For $B \parallel a$, the $\kappa(B)/\kappa(0)$ isotherms display a weak increase ($\sim$ 30 \%) at $T =$ 151 mK, while the magnetic-field dependence of $\kappa$ weakens gradually with increasing temperature, as shown in Fig. 5(a). In contrast, the $\kappa(B)/\kappa(0)$ is significantly enhanced at high magnetic fields for $B \parallel c$, as shown in Fig. 5(b). At $T =$ 151 mK, the $\kappa(B)/\kappa(0)$ quickly reaches saturation around $B =$ 1 T, which means at such low temperature the spins are easily polarized by the $c$-axis magnetic field. In other word, the proposed QSL state and spinon excitations can be easily suppressed by the $c$-axis magnetic field. As the temperature increases, the saturated magnetic field in $\kappa(B)/\kappa(0)$ curve gradually increases, which is also compatible with the magnetization behavior shown in Fig. 2(b). At $T =$ 5 K, the high-field enhancement of $\kappa(B)/\kappa(0)$ can be as high as 540 \%, which clearly reveals the strong scattering between magnetic excitations and phonons at zero field. The above significant different field dependence of $\kappa$ between $B \parallel c$ and $B \parallel a$ further demonstrates the strong spin anisotropy in PrMgAl$_{11}$O$_{19}$.

\section{DISCUSSION}

In the exploration of the classification and ground state characteristics of QSL candidates, disorder plays an important role. Due to the large difference in the ionic radius of Pr$^{3+}$ and Al$^{3+}$/Mg$^{2+}$ ions, the site mixing between magnetic and nonmagnetic ions is effectively avoided in PrMgAl$_{11}$O$_{19}$. But there is a weak site mixing between the nonmagnetic Al$^{3+}$ and Mg$^{2+}$ ions, as shown in Fig. 1(a), which is similar to the case of YbMgGaO$_4$, in which nonmagnetic Ga$^{3+}$ and Mg$^{2+}$ ions randomly distribute \cite{SciRep5, JAlloysCompd}. Since the occupation mixing only exists in Mg/AlO$_4$ tetrahedra with small proportion, the disorder in PrMgAl$_{11}$O$_{19}$ is much weaker compared with the case of YbMgGaO$_4$ \cite{JAlloysCompd, PhysRevB165143}. Moreover, the nearest-neighbor magnetic layers are separated by four layers of nonmagnetic Al/Mg-O polyhedrons, the disorder effect should be negligible for the ground-state properties and all of our measurements reflect the intrinsic properties of PrMgAl$_{11}$O$_{19}$. Our magnetic susceptibility results differ by two orders of magnitude for $B \parallel c$ and $B \parallel a$, indicating the Ising-like characteristics, which is consistent with the results reported recently \cite{PhysRevB165143}. In addition, the specific heat indicates the absence of long-range magnetic order and the power-law temperature dependence of magnetic specific heat, which points to a possible gapless QSL state. The spin dynamics results of PrMgAl$_{11}$O$_{19}$ are similar to its sister compound PrZnAl$_{11}$O$_{19}$, which has been claimed to possess a QSL with gapless spinon excitations \cite{PhysRevB134428}. Recently, the quasi-quadratic temperature-dependent behavior of magnetic specific heat has already been reported and a broad continuum of low-energy magnetic excitation spectra was captured by the inelastic neutron scattering technique at 55 mK, evidencing the possible realization of a gapless QSL state in PrMgAl$_{11}$O$_{19}$ \cite{PhysRevB165143}.

Ultralow-temperature thermal conductivity has been proved to be an effective probe for the itinerant spinons in the QSL candidates \cite{PhysRevB184423, Science328, NC4949, NC4216, NP44, PRX041051, PhysRevsearch013099}. After nearly twenty years studying on the ultralow-temperature thermal conductivity of QSL candidates, several typical results can be arrived:

(i) The $\kappa$ displays not only the $T^3$ phonon term but also an additional exponential term ($\propto$ exp($-\Delta/k_BT$)), which indicates the gapped QSL, as reported in the frustrated triangular magnet $\kappa$-(BEDT-TTF)$_2$Cu$_2$(CN)$_3$ \cite{NP44}.

(ii) The plot of $\kappa/T$ as a function of $T^2$ shows good linear behavior in a rather broad temperature range. This is the clearest case that the phonon thermal conductivity achieves the boundary scattering limit at low temperatures and exhibits a $T^3$ behavior. In this case, the non-zero residual term ($\kappa_0/T$) would strongly evidence the existence of itinerant gapless spinons. The most famous examples are EtMe$_3$Sb[Pd(dmit)$_2$]$_2$ \cite{Science328}, 1$T$-TaS$_2$ \cite{PhysRevsearch013099}, and Na$_2$BaCo(PO$_4$)$_2$ \cite{NC4216}. On the other hand, if the residual term is zero, one can definitely conclude the absence of spinons. It should be noted that, there is a well-known controversy on the results of some materials, since the zero residual term was also reported. For example, two groups observed zero residual term in EtMe$_3$Sb[Pd(dmit)$_2$]$_2$ \cite{PRX041051, PRL247204}. There could be two reasons for this experimental disagreement. First, as pointed out by Yamashita {\it et al.} based on systemic further measurements \cite{PRB140407}, the residual term is strongly affected by some scattering effect or sample imperfectness. Second, the scattering between spinons and phonons might also affect the residual term, as discussed below. This possibility is based on the experimental result that there is no linear relationship between $\kappa/T$ and $T^2$ in the data reported by Bourgeois-Hope {\it et al.} and Ni {\it et al.} \cite{PRX041051, PRL247204}, which is essentially different from the results reported by Yamashita {\it et al.} \cite{Science328}.

(iii) The plot of $\kappa/T$ as a function of $T^\alpha$ ($\alpha$ is much smaller than 2 or close to 1) shows good linear behavior at very low temperatures. In this case, the strong deviation of the phonon term from the standard phonon boundary-scattering limit ($T^3$-dependence) indicates rather strong scattering or coupling between the phonons and spinons. One may note that in some materials, the phonon reflection effect can give a temperature dependence of phonon thermal conductivity weaker than $T^3$; but with this effect the temperature dependence is still rather close to $T^3$ \cite{PRB104412, PRB081111}. Here, one necessary experimental signature to verify the spinon-phonon scattering is the enhancement of thermal conductivity under high magnetic fields. When the above two phenomena (very weak $T$-dependence of $\kappa$ and the high-field enhancement of $\kappa$) are simultaneously observed in QSL candidates, one can conclude the existence of gapless spinons, which can both transport heat and scatter with phonons. Among these materials, the non-zero residual term would also evidence the existence of itinerant gapless spinons. The examples are YbMgGaO$_4$ \cite{NC4949} and Na$_2$Co$_2$TeO$_6$ \cite{PhysRevB184423}, even though the ability of spinon transport is rather weak because of the scattering effect. In comparison, PrMgAl$_{11}$O$_{19}$ actually displays an extreme example that the spinon-phonon scattering is so strong that we can only observe an almost zero $\kappa_0/T$, as well as the linear function of $\kappa/T$ as $T$ at $B =$ 0 T; in addition, the effect of applying magnetic field on the thermal conductivity verifies such scattering.

Finally, it should be ponited out that the spin-phonon coupling is common in various types of materials and in many cases the magnetic field suppression of the coupling on the results of thermal conductivity may not be relevant to the spinon excitations. However, for PrMgAl$_{11}$O$_{19}$, the previous specific heat and neutron scattering results, as well as our own specific heat data, have already suggested the gapless QSL ground state. Thus, the observed linear function of $\kappa/T$ as $T$ at $B =$ 0 T and the magnetic field enhanced thermal conductivity can represent the spinon-phonon scattering. One advantage of using thermal conductivity is that at very low temperatures such disordered spin systems (no long-range order) cannot have other magnetic excitations than spinons that yield strong scattering effect on phonons; furthermore, such scattering effect is monotonically suppressed with increasing field. In contrast, the localized spins can also scatter phonons but will yield non-monotonic field dependence of thermal conductivity \cite{Oxford}.

Therefore, based on the previous studies and summarization all the possibilities, the ultralow-temperature thermal conductivity of PrMgAl$_{11}$O$_{19}$ can demonstrate the existence of gapless spinons that strongly scatter with phonons. One aim of this work is to show this particular experimental probe for the gapless spinons. It is likely that similar phenomenon will be observed in some other QSL candidates. Another peculiar phenomenon found in this work is that the $a$-axis magnetic field affects the $\kappa$ very weakly, in contrast to the case of applying $c$-axis magnetic field. As discussed above, this can be understood based on the Ising-like anisotropy of PrMgAl$_{11}$O$_{19}$. At present, the Ising-type QSL candidates are very rare and the relevant physics needs further investigations.

\section{SUMMARY}

We have successfully grown centimeter-sized single crystal of a perfect triangular-lattice compound PrMgAl$_{11}$O$_{19}$ by using the optical floating-zone technique. The magnetic susceptibility reveals Ising-like interactions with antiferromagnetic correlations between Pr$^{3+}$ ions. Both specific heat and thermal conductivity measurements reveal the absence of long-range magnetic order down to several tens of milliKelvin. Furthermore, the zero-field $C_p(T)$ curve exhibits power-law $T$-dependent behavior at low temperatures, reminiscent of the gapless magnetic excitations in QSL. The temperature and magnetic field dependencies of ultralow-temperature thermal conductivity suggest that there are gapless spinons scattering with phonons rather than transporting heat.

\begin{acknowledgements}

This work was supported by the National Key Research and Development Program of China (Grant No. 2023YFA1406500), the National Natural Science Foundation of China (Grant Nos. 12274388, 12104010, 12104011, and 12075205) and the Nature Science Foundation of Anhui Province (Grant Nos. 1908085MA09 and 2108085QA22). The work at the University of Tennessee was supported by the NSF with Grant No. NSF-DMR-2003117. H.W. and W.X. are supported by the U.S. DOE-BES under Contract No. DE-SC0023648.

\end{acknowledgements}

\end{document}